\documentclass[runningheads,a4paper]{llncs}

\usepackage{amssymb}
\usepackage{graphicx}

\usepackage{etoolbox}
\usepackage[skip=0pt]{caption}
\usepackage{fixltx2e}
\usepackage{caption}
\usepackage{amsmath}
\usepackage{graphicx}
\usepackage{multirow}
\usepackage{amssymb}
\usepackage{color}
\usepackage{graphicx}
\usepackage{url}
\usepackage{enumerate}
\usepackage{balance}
\usepackage{array}
\usepackage{appendix}
\mathchardef\mhyphen="2D

\usepackage{url}
\urldef{\mailsa}\path|{roylee.2013@smu.edu.sg|
\urldef{\mailsb}\path|davidlo@smu.edu.sg}|
\newcommand{\keywords}[1]{\par\addvspace\baselineskip
\noindent\keywordname\enspace\ignorespaces#1}

\begin{document}

\mainmatter  % start of an individual contribution

\title{GitHub and Stack Overflow: Analyzing Developer Interests Across Multiple Social Collaborative Platforms}

\titlerunning{Developer Interests Across Multiple Social Collaborative Platforms}

\author{Roy Ka-Wei Lee
\and David Lo}
\authorrunning{Developer Interests Across Multiple Social Collaborative Platforms}

\institute{School of Information Systems,\\
Singapore Management University\\
\mailsa\\
\mailsb\\}


\toctitle{Lecture Notes in Computer Science}
\tocauthor{Authors' Instructions}
\maketitle

\begin{abstract}
Increasingly, software developers are using a wide array of social collaborative platforms for software development and learning. In this work, we examined the similarities in developer's interests within and across GitHub and Stack Overflow. Our study finds that developers share common interests in GitHub and Stack Overflow; on average, 39\% of the GitHub repositories and Stack Overflow questions that a developer had participated fall in the common interests. Also, developers do share similar interests with other developers who co-participated activities in the two platforms. In particular, developers who co-commit and co-pull-request same GitHub repositories and co-answer same Stack Overflow questions, share more common interests compare to other developers who co-participate in other platform activities.  
\end{abstract}

\section{Introduction}
\label{sec:introduction}
Software developers are increasingly adopting social collaborative platforms for software development and making a reputation for themselves. Two of such widely adopted and studied social-collaborative platforms are \textit{GitHub}\footnote{https://github.com/} and \textit{Stack Overflow}\footnote{http://stackoverflow.com/}. GitHub is a collaborative software development platform that allows code sharing and version control. Developers can participate in various activities in GitHub, for example, developers may \textit{fork} (i.e., create a copy of) repositories of other developers. Stack Overflow is a community-based website for asking and answering questions relating to programming languages, software engineering, and tools. Although the two platforms are used for different purposes, developers can utilize both platforms for software development. For example, a developer who has interests in Java programming language may fork a Java project in GitHub and answer Java programming questions in Stack Overflow.

We broadly define the interests of a developer as the programming related topic domains of GitHub repositories and Stack Overflow questions that he or she has participated. For instance, when a developer answers questions tagged with {\em javascript}, {\em jquery}, and {\em angularjs}, we deduce that the developer is interested in the three technologies. Similarly, when a developer forked repositories in GitHub which description contains keywords such as {\em javascript} and {\em ajax}, we could estimate that the developer is interested in the two technologies.

The learning of developers' interests could provide new insights to how developers utilize the two social collaborative platforms for software development. For example, if developers share similar interests in GitHub and Stack Overflow, the two platforms may be used to complement each other for software development. Conversely, if the developers display differences between their interests in GitHub and Stack Overflow, the two platforms may have been used in a disjoint manner. The social and community-based element in GitHub and Stack Overflow also adds on to the dynamics when studying developer's interests; developers may find themselves sharing similar interests with other developers who also co-participated in a common repository or question. Thus, it would be interesting to investigate the interests of developers within and across the two platforms. In particular, we ask the following research questions: Does an individual developer share similar interests in his GitHub and Stack Overflow accounts? (\textbf{RQ1}), and does an individual developer share similar interests with other developers who co-participated activities in GitHub and Stack Overflow? (\textbf{RQ2}).

Our research in this paper is thus divided into two main parts: In the first part, we propose similarity scores to measure the developer's interests within and across social collaborative platforms. In the second part, we applied the propose measures on large GitHub and Stack Overflow datasets and conduct an empirical study to answer the two research questions listed earlier. 

\textbf{Contributions}. This work improves the state-of-the-art of inter-network studies on multiple social collaborative platforms. Key contributions of this work include: Firstly, to the best of our knowledge, it is the first research attempt to study similarity of developer interests across GitHub and Stack Overflow using large datasets. Second, we proposed several scores to measure the similarity in developer interests within and across social collaborative platforms. The proposed similarity scores are also applied in an empirical study to quantify the similarity in developer's interests within and across Stack Overflow and GitHub.  

%========================================================================================
\section{Data Preparation}
\label{sec:data}

\subsection{Dataset}
There are two main datasets used in our study; we retrieve activities from October 2013 to March 2015 of about 2.5 million GitHub users and 1 million Stack Overflow users from open-source database dumps\cite{Gousi13}\footnote{https://archive.org/details/stackexchange}. As this study intends to investigate developer interests across GitHub and Stack Overflow, we further identify developers who were using both platforms. For this work, we used the dataset provided by Badashian et al. \cite{badashian2014}, where they utilized GitHub users' email addresses and Stack Overflow users' email MD5 hashes to find the intersection between the two datasets. In total, we identify 92,427 developers, which forms our \textit{base developer} set. Subsequently, we retrieved the platform activities participated by the base developers. In total, we have extracted 416,171 \textit{Fork}, 2,168,871 \textit{Watch},  846,862 \textit{Commit}, 386,578 \textit{Pull-Request}, 277,346 \textit{Ask}, 766,315 \textit{Answer} and 427,093 \textit{Favorite} activities. Our subsequent analysis will be based on this group of activities participated by the base developers.

\subsection{Estimating Developer Interests}\label{subsec:devint}

We estimate developer interests by observing the group of activities they participated in GitHub and Stack Overflow. To estimate developer interests in Stack Overflow, we use the descriptive tags of the questions that they asked, answered and favorited. For example, consider a question $q$ related to mobile programming for Android smartphones which contain the following set of descriptive tags: \{\textit{Java, Android}\}. If a developer $d$ asked, answered, or favorited that question, we estimate that his interests include \textit{Java} and \textit{Android}. GitHub does not allow users to tag repositories but it allows users to describe their repositories. These descriptions often contain important keywords that can shed light to developer interests. To estimate developer interests from the repositories that a developer had participated, we first collect all descriptive tags that appear in our Stack Overflow dataset. Subsequently, we perform keyword matching between the collected Stack Overflow tags and a GitHub repository description. We consider the matched keywords as the estimated interests. We choose to use Stack Overflow tags to ensure that developer interests across the two platforms can be mapped to the same vocabulary.

We denote the estimated interests of a developer given a repository $r$ that he or she forked, watched, committed or pull-requested in GitHub as $I(r)$. Similarly, we denote the estimated interests of a developer given a question $q$ that he or she asked, answered, or favorited in Stack Overflow as $I(q)$. Since the estimated interests given a repository or a question is the same for all developers participated in it, we also refer to $I(r)$ and $I(q)$ as the interests in $r$ and $q$. For simplicity, we also refer to them as $r$'s interests and $q$'s interests respectively. Developer $d$'s overall interest in GitHub and Stack Overflow, denoted by $I^{GH}(d)$ and $I^{SO}(d)$, is the union of his/her interests over all the repositories and questions group of activities that $d$ has participated in.

%========================================================================================
\section{Measuring Developer Interests Similarity}
\label{sec:measures}

\subsection{Developer Interests Similarity Across Platforms}
One way to measure the similarity in an individual developer's interests across platforms is to take the intersection of his interests in Stack Overflow ($I^{SO}(d)$) and his interests in GitHub ($I^{SO}(d)$). However, this simple measure considers all interests to have an equal weight. In reality, a developer may ask much more questions related to a particular interest than other interests. Similarly, a developer may fork repositories related to a particular interest than other interests. Thus, a finer way to measure the similarity in developer interests should consider the number of repositories and questions that belong to each interest.

To capture the above mentioned intuition, we propose \textit{cross-platform similarity score}, which is denoted as $\mathit{Sim^{SO\mhyphen GH}(d)}$. Given a developer $d$, we measure $d$'s similarity in interests across Stack Overflow (SO) and GitHub (GH) by computing the {\em proportion} of $d$'s repositories and questions that fall in $d$'s {\em common interests} in Stack Overflow and GitHub (i.e., $I^{SO}(d)$ $\bigcap$ $I^{GH}(d)$). By denoting the repositories and questions that are related to $d$ (i.e., $d$ forked, watched, committed, pull-requested, asked, answered, or favorited these repositories or questions) as $d.R$ and $d.Q$, we can mathematically define $\mathit{Sim^{SO\mhyphen GH}(d)}$ as follows:

\begin{equation} \label{sogh_eqn4}
CI(d) = I^{SO}(d) \bigcap I^{GH}(d)
\end{equation}

\begin{equation} \label{sogh_eqn3}
Shared^{Q}(d) = \{q\in d.Q|I(q) \in CI(d)\}\\
\end{equation}

\begin{equation} \label{sogh_eqn2}
Shared^{R}(d) = \{r\in d.R|I(r) \in CI(d)\}\\
\end{equation}

\begin{equation} \label{sogh_eqn}
Sim^{SO\mhyphen GH}(d) = \frac{|Shared^{R}(d)| + |Shared^{Q}(d)|}{|d.R|+|d.Q|}
\end{equation}

In Equation~\ref{sogh_eqn4}, we define the common interests of developer $d$ in both Stack Overflow and GitHub. Equation~\ref{sogh_eqn3} defines the set of questions that falls into the common interests, while Equation~\ref{sogh_eqn2} defines the set of repositories that falls into the common interests. Equation~\ref{sogh_eqn} defines $\mathit{Sim^{SO\mhyphen GH}(d)}$ as the proportion of repositories and questions of $d$ that falls into the common interests. Please refer to Appendix 1 for an example that illustrate how \textit{cross-platform similarity score} is calculated.

\subsection{Developer Interests Similarity Among Co-Participated Developers}

To study the similarity of interests among developers who co-participated in GitHub and Stack Overflow activities, we propose \textit{co-participation similarity scores}, each focusing on a platform activity. Given a platform activity and a target developer $d$, we want to measure the similarity between $d$ and {\em all other developers} who co-participated in the target activity for {\em at least one} common GitHub repository or StackOverflow question. For example, considering forking a repository as an activity of interest, we want to find developers who co-fork at least one common GitHub repository with $d$. Hence, given a developer $d$, we denote the set of other developers who co-participated in forking at least one common repository or question as $Co^F(d)$.

Intuitively, the more repositories or questions of common interests that $d$ share with other developers in $Co^F(d)$, the higher the similarities should be. To compute the similarity in interests between $d$ and $Co^F(d)$, we measure the average similarity in interests between $d$ and each developer $d'$ in $Co^F(d)$; for each of such pair, we measure their similarity by computing the proportion of $d'$'s forked repositories which share an interest with the interests of $d$ in his/her forked repositories.  Mathematically, we define the \textit{co-participation similarity scores} for forking in Equation~\ref{cofork_eqn}.

\thickmuskip=0.1\thickmuskip
\thinmuskip=0.1\thinmuskip
\medmuskip=0.1\medmuskip

\begin{equation} \label{cofork_eqn}
Sim^F(d,Co^F(d)) = \frac{\sum_{d'\in Co^F(d)} \frac{|Shared^F(d,d')|}{|d'.RF|}}{|Co^F(d)|}
\end{equation}

\begin{equation} \label{cowatch_eqn}
Sim^W(d,Co^W(d)) = \frac{\sum_{d'\in Co^W(d)} \frac{|Shared^W(d,d')|}{|d'.RW|}}{|Co^W(d)|}
\end{equation}

\begin{equation} \label{cocommit_eqn}
Sim^C(d,Co^C(d)) = \frac{\sum_{d'\in Co^C(d)} \frac{|Shared^C(d,d')|}{|d'.RC|}}{|Co^C(d)|}
\end{equation}

\begin{equation} \label{copullrequest_eqn}
Sim^P(d,Co^P(d)) = \frac{\sum_{d'\in Co^P(d)} \frac{|Shared^P(d,d')|}{|d'.RP|}}{|Co^P(d)|}
\end{equation}

\begin{equation} \label{coanswer_eqn}
Sim^A(d,Co^A(d)) = \frac{\sum_{d'\in Co^A(d)} \frac{|Shared^A(d,d')|}{|d'.QA|}}{|Co^A(d)|}
\end{equation}

\begin{equation} \label{cofavorite_eqn}
Sim^V(d,Co^V(d)) = \frac{\sum_{d'\in Co^V(d)} \frac{|Shared^V(d,d')|}{|d'.QV|}}{|Co^V(d)|}
\end{equation}
 
In the above formulas, $d'.RF$ denotes the repositories or questions that $d'$ forked. Furthermore, $\mathit{Shared^F}$ $\mathit{(d,d')}$ denotes the set of repositories which are forked by $d'$ and share common interests with $d$'s forked repositories. Mathematically, it is defined as:

\begin{equation} \label{share_eq}
\{r'\in d'.RF|\bigg[I(r') \bigcap \bigcup_{r\in d.RF} I(r)\bigg] \neq \emptyset\}
\nonumber
\end{equation}

In Equation \ref{cofork_eqn}, we define the average similarity in interests between developer \textit{d} and other developers who had co-forked at least 1 repository with \textit{d}. The \textit{co-participation similarity scores} for co-watch ($Sim^W(d,Co^W(d))$), co-commit ($Sim^C(d,Co^C(d))$), co-pull-request ($Sim^P(d,Co^P(d))$), co-answer \\($Sim^A(d,Co^A(d))$), and co-favorite ($Sim^V(d,Co^V(d))$) are similarly defined in Equation \ref{cowatch_eqn} to \ref{cofavorite_eqn}. Please refer to Appendix 2 for an example that illustrate how \textit{co-participation similarity score} is calculated.

%========================================================================================
\section{Empirical Study}
\label{sec:empirical}
In this section, we applied the developer interests similarity measures proposed in the previous section on GitHub and Stack Overflow large datasets. We also attempt to answer the two research questions that we have listed earlier in this empirical study \textbf{ RQ1} and \textbf{RQ2}.

\subsection{RQ1: Does an individual developer share similar interests in his GitHub and Stack Overflow account?}
\begin{figure} [h]
	\begin{center}
		\includegraphics[scale = 0.32]{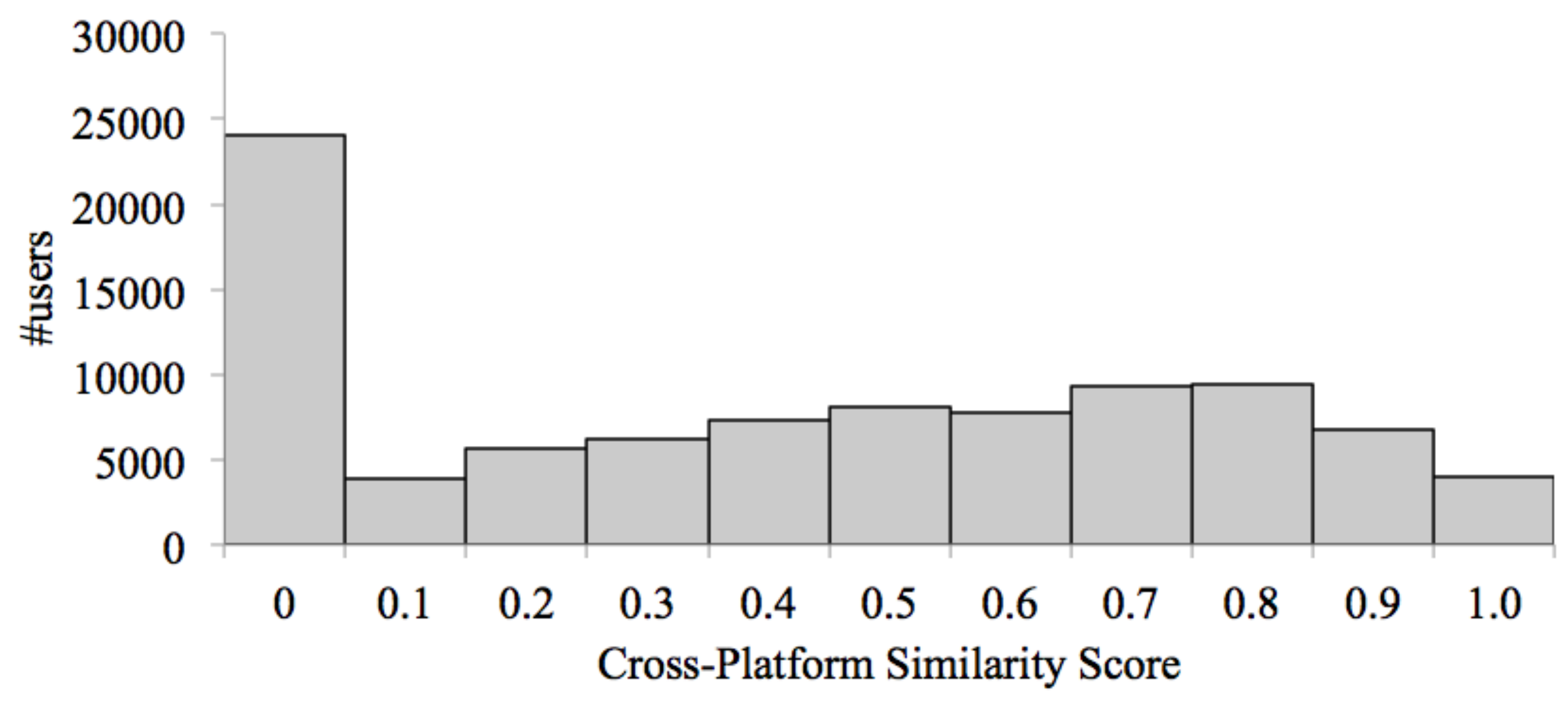}
		\caption{Distribution of developers' \textit{cross-platform similarity scores} in GitHub and Stack Overflow}
		\label{fig:simcrossdist}
	\end{center}
\end{figure}

Figure \ref{fig:simcrossdist} shows the distribution of the \textit{cross-platform similarity scores} computed for the base developers. On average, the developers have a similarity score of 0.39. This suggests that on average, 39\% of the GitHub repositories and Stack Overflow questions that a developer had participated shared similar interests. Also, close to half (49\%) of the developers have scored 0.5 or higher, while 26\% of the developers have their similarity scores equal to 0, i.e., the interests of these developers are totally different in GitHub and Stack Overflow. This suggests that although most developers do share high similarity in interests in GitHub and Stack Overflow, however, there are a group of developers who have totally different interest in GitHub and Stack Overflow. 

\vspace{-6mm}

We further drill down to compare the similarity in developer interests for different types of activity across the two platforms. For example, we measure the similarity in developer's interests by only considering repositories that the developer has forked and questions that the developer has answered. Twelve different combinations capturing different pairs of activities across the two platforms are considered: \textit{Fork-Ask}, \textit{Fork-Answer}, \textit{Fork-Favorite}, \textit{Commit-Ask}, \textit{Commit-Answer}, \textit{Commit-Favorite}, \textit{pull-request-Ask}, \textit{pull-request-Answer}, \textit{pull-request-Favorite}, \textit{Watch-Ask}, \textit{Watch-Answer} and \textit{Watch-Favorite}.

\begin{figure} [h]
	\begin{center}
		\includegraphics[scale = 0.42]{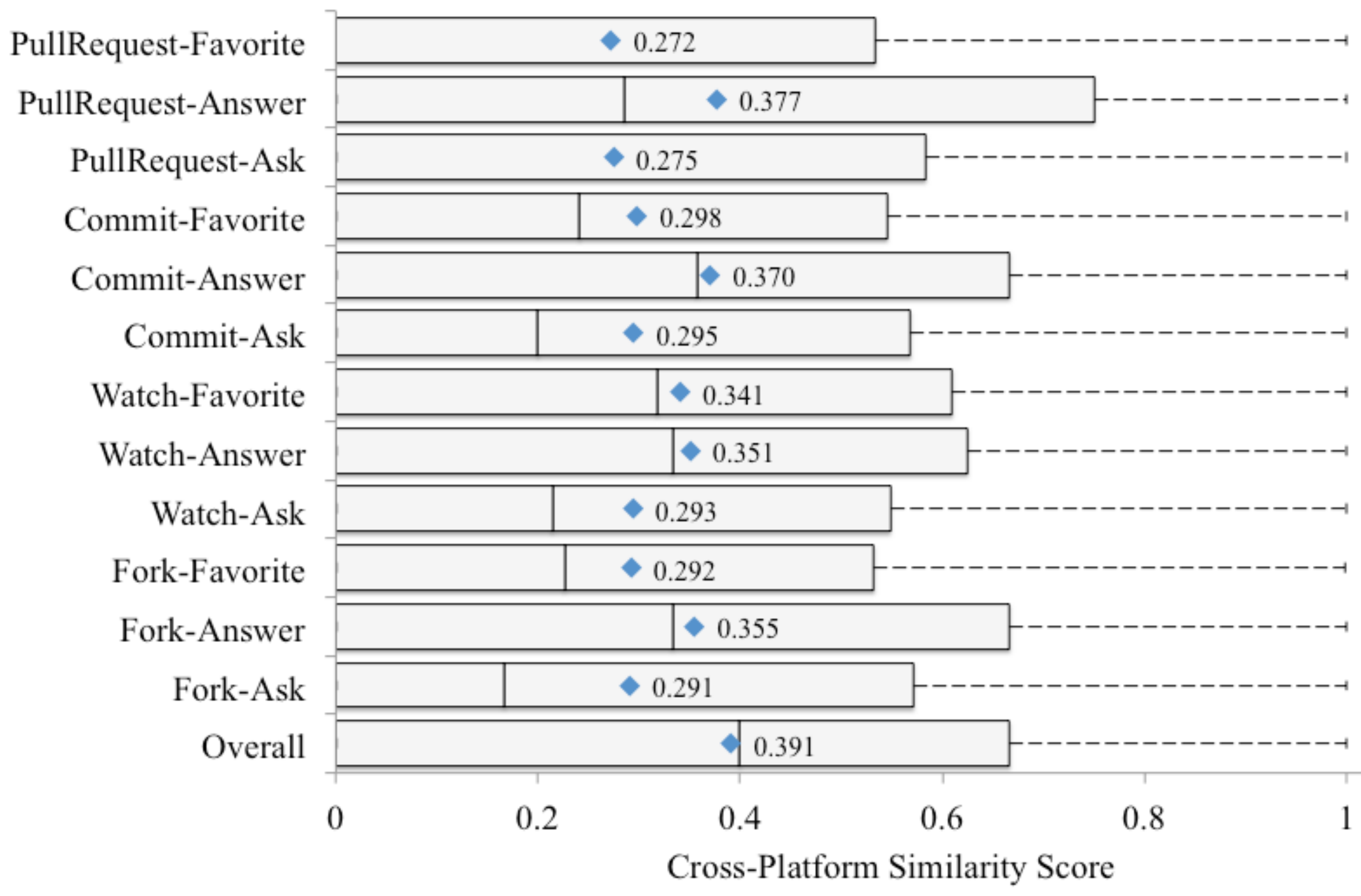}
		\caption{Boxplots of interest similarity for different activity pairs}
		\label{fig:boxplotcross}
	\end{center}
\end{figure}

Figure \ref{fig:boxplotcross} shows the boxplots of \textit{cross-platform similarity scores} for the 12 different activity pairs. The platform activity pairs have average similarity scores between 0.27 to 0.38, slightly lower than the overall average of 0.39. All the platform activity pairs also have significantly higher number of developers with scores of 0. This is as expected since by combining all platform activity pairs we have a larger pool of common interests. Among the 12 activity pairs, \textit{pull-request-Answer} pair has the highest average similarity score. A possible explanation for this observation could be attributed to the nature of the platform activity; \textit{pull-request} and \textit{answer} not only reveal the interests of the developers but also demand the developers to have a certain expertise on the topics or programming languages of the participated repositories and questions. For example, a developer who is proficient in Java programming language would only \textit{answer} Java programming related questions and submit \textit{pull-request} for Java repositories but he could \textit{watch} other programming language repositories or \textit{favorite} questions from other topics for learning purposes.

\vspace{-10mm}

\subsection{RQ2: Does an individual developer share similar interests with other developers who co-participated activities in GitHub and Stack Overflow?}

\begin{figure}[h]
	\begin{center}
		\includegraphics[scale = 0.37]{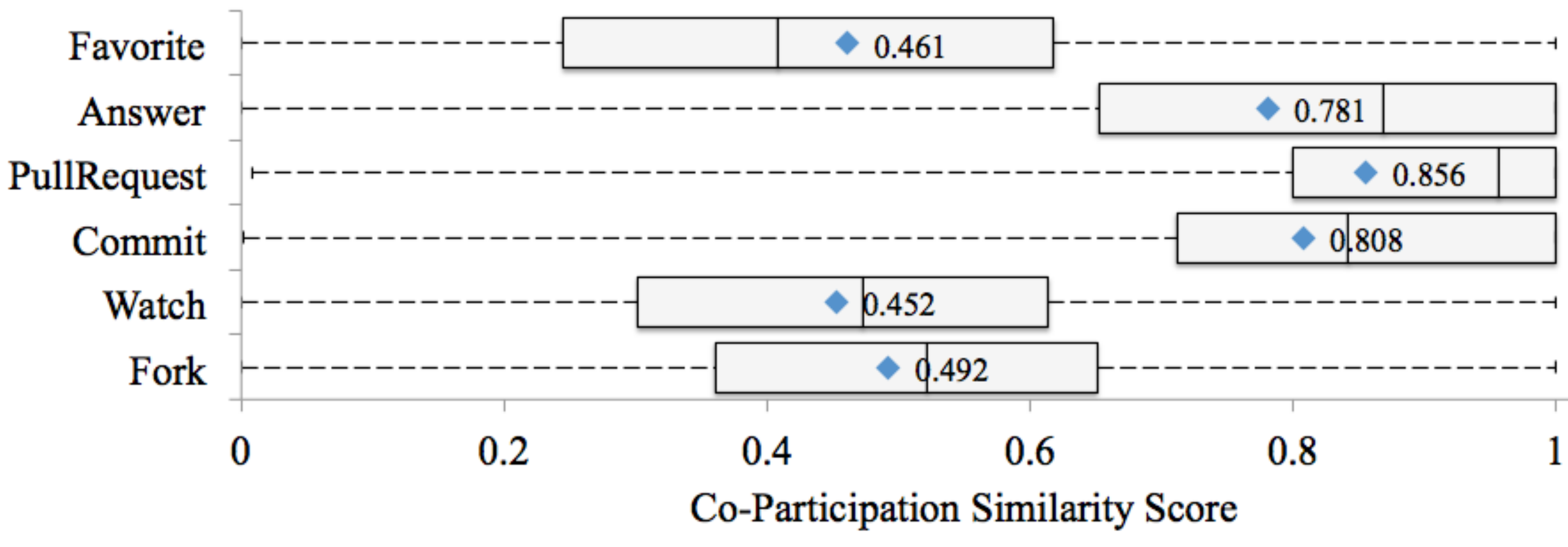}
		\caption{Boxplots of \textit{Co-Participation Similarity} scores for different activities}
		\label{fig:boxplotneighbour}
	\end{center}
\end{figure}

Figure \ref{fig:boxplotneighbour} shows the boxplots of \textit{co-participation similarity scores} of the base developers. We observe that an individual developer has average similarity scores between 0.45 to 0.86 with other developers who participated in at least one common platform activity. This means that given two developers who participated in a common platform activity, on average 45-86\% of all repositories and questions that they participated in that platform activity shared common interests. Interestingly, we also observed that \textit{commit}, \textit{pull-request} and \textit{answer} have higher average similarity score compare to the rest of the platform activities (0.81, 0.86 and 0.78 respectively). A possible reason for this observation could again be related to the expertise of the developers. We would expect that the expertise of the developers to be more specialized and less diverse than developers' interests, thus resulting in higher similarity scores for developers sharing a common \textit{commit}, \textit{pull-request} and \textit{answer}. 

%\vspace{-4mm}
\vspace{-10mm}

\subsection{Discussion}
Our empirical study has validated that developers do display some similarity in interests in their  GitHub and Stack Overflow accounts (\textbf{RQ1}) and developers do share common interests with other co-participating developers in the platforms (\textbf{RQ2}). Furthermore, we were able to quantify the level of similarity in developer interests across different social collaborative platforms; we found that on average, 39\% of the GitHub repositories and Stack Overflow questions that a developer had participated fall in the common interests. The findings in this research could also spark more inter-platforms software engineering research. For instance, when studying the evolution of developer interests, one could take a different perspective and investigate the differences in developer interests in multiple social collaborative platforms over time to observe how developers learn and pick up new interests (e.g., a new programming language).

The findings from our empirical study could be extended to build predictive analytics and recommendation application. As we learned that developers do share interest similarity across platforms (\textbf{RQ1}), intuitively we could predict a developer's activities in one platform using his or her interests displayed on another platform. For example, if we learn that a developer answer Java related questions in Stack Overflow, and he displays high similarity in interests across platforms, we can predict that the developer is likely to participate in Java related repositories in GitHub. Likewise in our empirical study, we found that developers do share similar interests with other developers who co-participated activities in GitHub and Stack Overflow (\textbf{RQ2}). With this insights, we could predict a developer' activities in a platform using the interests of other developers who had co-participated with him or her in the platform. For example, if we learn that a developer answers a Java related question in Stack Overflow, and we learn that other developers who answered the same questions also display strong interests in Android related questions, we can predict that the developer too, is likely to participate in Android related questions in Stack Overflow. We will look into extending our empirical study predictive analytics and recommendation systems in future works.

%========================================================================================
\section{Related Work}
\label{sec:related}
There have been few existing inter-network studies on GitHub and Stack Overflow. These works did deepen our understanding of developer behaviors in the two social-collaborative platforms. Vasilescu et al. performed a study on developers' involvement and productivity in Stack Overflow and GitHub~\cite{vasilescu2013}. They found that developers who are more active on GitHub (in terms of GitHub commits), tend to ask and answer more questions on Stack Overflow. Badashian et al. \cite{badashian2014} did an empirical study on the correlation between different types of developer activities in the two platforms. Their findings supported the findings of the earlier work by Vasilescu et al., that is: developers who actively contributed to GitHub, also actively answered questions in Stack Overflow. They observed overall weak correlation between the activity metrics of the two networks and concluded that developer activities in one network are not strong predictors for activities on another network. Both the works, however, did not consider intrinsic interests of the developers, although Vasilescu et al. did mention the possibility of extending their work to consider topic interests of the developers. 

Stack Overflow and GitHub have also been studied for empirical works on developer interests. For example, there were research works that focused on analyzing topics asked by developers in Stack Overflow~\cite{Barua2014,Bajaj2014,zou2015,rosen2015,yang2016security,wang2013}. Similarly, there were also works on analyzing programming languages used by developers in GitHub and their relationships to GitHub contributions~\cite{bissyande2013,Ray2014,Sheoran2014,vasilescu2015,Rahman2014,jiang2017,kochhar2017}. There are also studies characterizing social network properties of GitHub and Stack Overflow~\cite{thung2013,wang2013}. Our work extends this group of research by comparing developer interests in the two social collaborative platforms. To our best of knowledge, our work is the first inter-network study that examines cross-site developer interests in GitHub and Stack Overflow.

%========================================================================================
\section{Conclusion and Future Work}
\label{sec:conclusion}
In this paper, we studied the similarity in developer interests within and across GitHub and Stack Overflow. Our findings were based on data for 92,427 users who were active in GitHub and Stack Overflow. We first proposed similarity scores to measure similarity in developers' interests within and across social collaborative platforms. Next, we applied our proposed similarity scores in an empirical study on GitHub and Stack Overflow. We observed that on average, 39\% of the GitHub repositories and Stack Overflow questions that a developer had participated fall in the common interests. The developers also do share common interests with other developers who co-participated activities in the platforms. For future works, we intend to we conducted experiments to predict the GitHub and Stack Overflow activities of developers using the insights gathered from our empirical analysis. For example, we can predict developer's GitHub activities using the interests learnt from his or her Stack Overflow activities, and vice versa. We also plan to conduct empirical studies to separate the expertises and interests of developers. 

\section*{Acknowledgments}
This research is supported by the National Research Foundation, Prime Minister's Office, Singapore under its International Research Centres in Singapore Funding Initiative. 

\section*{Appendix 1: Example for Cross-Platform Similarity Score Calculation}
\label{app:excrossplatform}

\begin{figure}[h]
	\begin{center}
		\includegraphics[scale = 0.35]{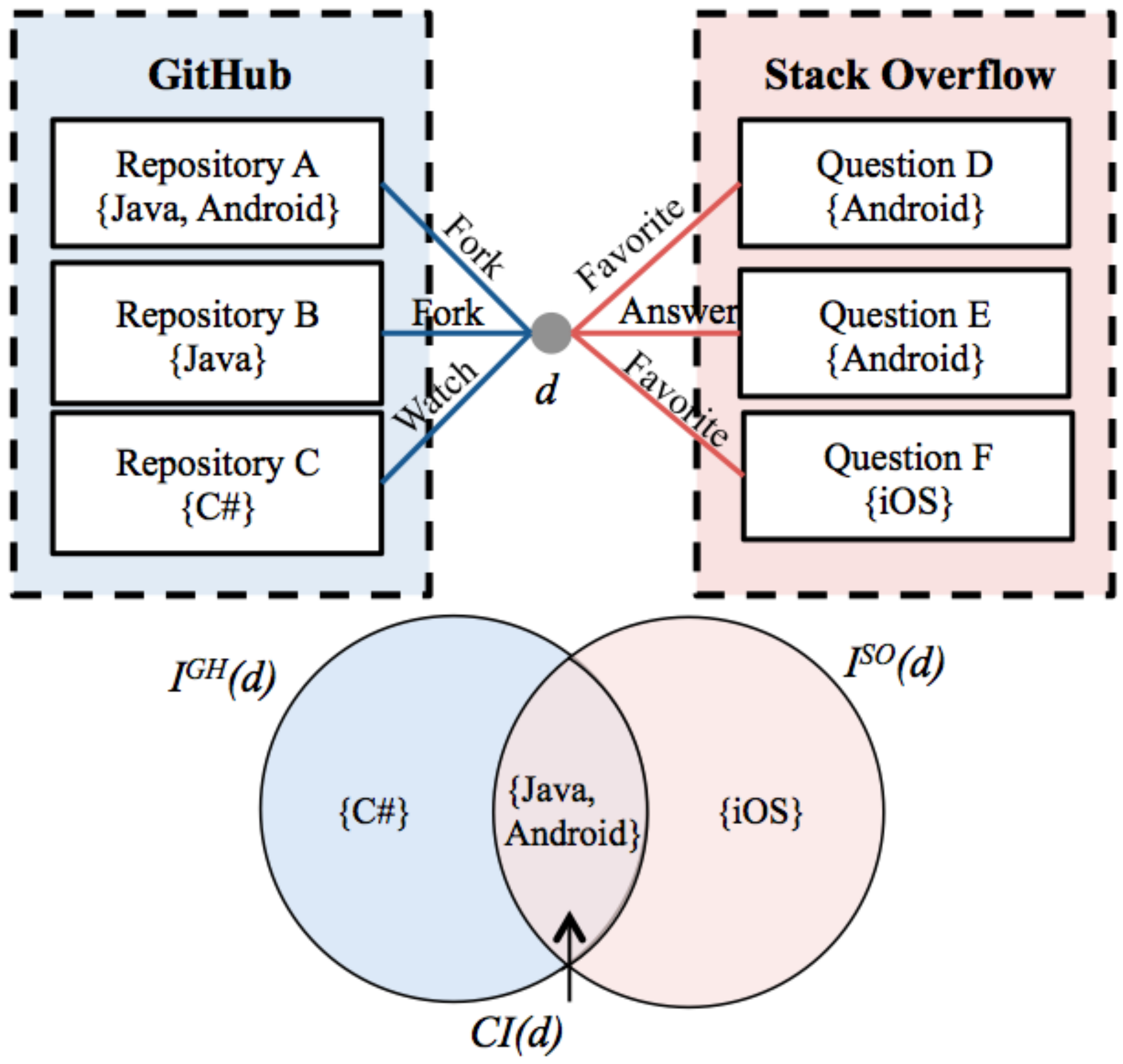}
		\caption{Example of \textit{cross-platform similarity score} calculation}
		\label{fig:soghexample}
	\end{center}
\end{figure}
\vspace{-6mm}

Figure \ref{fig:soghexample} shows an example for the calculation of \textit{cross-platform similarity score} $Sim^{SO\mhyphen GH}(d)$. Consider developer $d$ who has participated activities in GitHub and Stack Overflow. $d$ has forked 2 repositories; \textit{Repository A} which description contains the tag set \textit{\{Java, Android\}}, and \textit{Repository B} which description contains the tag set \textit{\{Java\}}, and watched \textit{Repository C} which description contains the tag set \textit{\{C\#\}}. $d$ also favorited 2 Stack Overflow questions; \textit{Question D} which are tagged with \textit{\{Android\}}, and \textit{Question F} which are tagged with \textit{\{iOS\}}, and answered \textit{Question E} which are tagged with \textit{\{Java\}}. We can estimate $d$'s interests in GitHub (i.e. $I^{GH}(d)$) as \textit{\{Java, Android, C\#\}} and $d$'s interests in Stack Overflow (i.e., $I^{SO}(d)$) as \textit{\{Android, iOS\}}. The common interests of $d$ (i.e., $CI(d)$) would be \textit{\{Java, Android\}}. Therefore, $Shared^{R}(d)$  would include repositories \textit{A} and \textit{B}, while $Shared^{Q}(d)$  would include questions \textit{D} and \textit{E}. Thus, $Sim^{SO\mhyphen GH}(d) = \frac{|2| + |2)|}{|3| + |3|}$.

\subsection*{Appendix 2: Example for Co-Participation Similarity Score Calculation}
\label{app:excoparticipation}

\begin{figure} [h]
	\begin{center}
		\includegraphics[scale = 0.3]{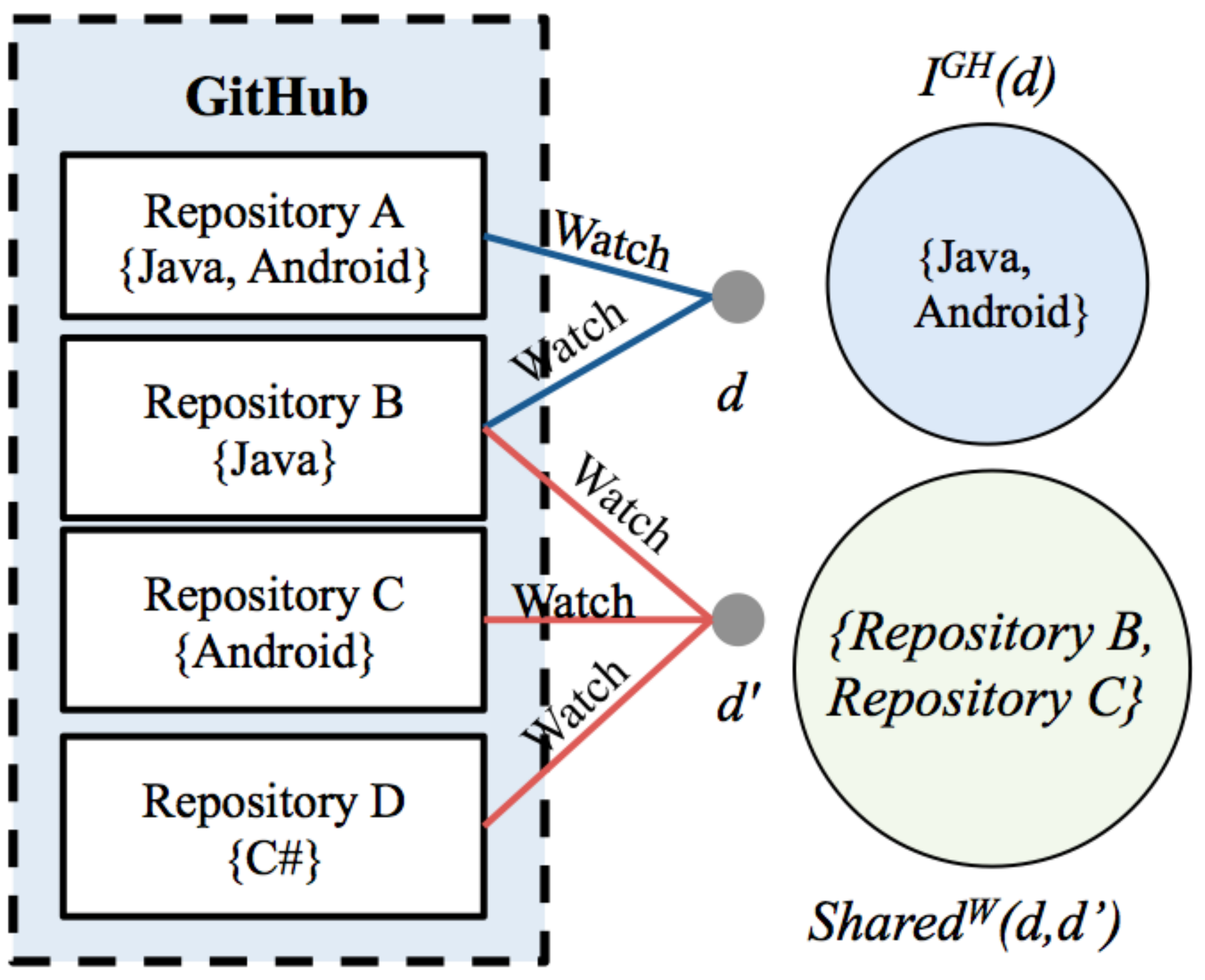}
		\caption{Example of \textit{co-participation similarity score} calculation for \textit{watch} activity}
		\label{fig:simcowatchexample}
	\end{center}
\end{figure}
\vspace{-6mm}

Figure \ref{fig:simcowatchexample} shows an example for the calculation of \textit{co-participation similarity score} for watch activity $Sim^W(d,co^W(d))$ for developer $d$. Let us consider two developers $d$ and $d'$ and assume that there are no other developers. Developer $d$ watched repositories \textit{ A} and \textit{B}. Developer $d'$ co-watched \textit{B} with $d$. Thus, $co^W(d)$ is \{d'\}. In addition to \textit{B}, developer $d'$ also watched repositories \textit{C} and \textit{D}. $Shared^W(d,d')$ would then include \textit{B} and \textit{C} as both of the repositories share common interests with the repositories that $d$ watched. $Sim^W(d) = \bigg[\sum_{d'\in Co^W(d)} \frac{|2|}{|3|}\bigg] / |1| = 0.67$.

It is important to note that the \textit{co-participation similarity scores} only consider the similarity in interests between pairs of developers who have co-participated in at least one common repository or question with each other but the developers may have participated in many other repositories and questions different from each other. For example, developers $d$ and $d'$ only watched one common repository but they had watched many other repositories which were different from each other. Also, when computing the co-participation similarity measure between developers who participated in a particular activity, we only consider the interests of the developers in that target activity. For instance, when computing $Sim^W(d)$ , we consider how similar are the interests between developers based only on the \textit{watch} activities, i.e., we do not consider repositories forked by the developers or questions answered and favorited by the developers. 

\bibliographystyle{plain}
\bibliography{ref}

\end{document}